\input{aipcheck}

\documentclass[
    ,final            
  ]
  {aipproc}

\layoutstyle{6x9}

\usepackage{graphicx}
\usepackage{amsmath}
\usepackage{latexsym}
\usepackage{amssymb}
\usepackage{dsfont}

\newcommand{\ud}{\mathrm{d}}

\begin{document}

\title{Accessing the quark orbital angular momentum with Wigner distributions}


\author{C\'edric Lorc\'e}{
  address={IPNO, Universit\'e Paris-Sud, CNRS/IN2P3, 91406 Orsay, France\\
and LPT, Universit\'e Paris-Sud, CNRS, 91406 Orsay, France\\
lorce@ipno.in2p3.fr}
}

\author{Barbara Pasquini}{
  address={Dipartimento di Fisica, Universit\`a degli Studi di Pavia, Pavia, Italy\\
and Istituto Nazionale di Fisica Nucleare, Sezione di Pavia, Pavia, Italy\\
Barbara.Pasquini@pv.infn.it}
}

\begin{abstract}
The quark orbital angular momentum (OAM) has been recognized as an important piece of the proton spin puzzle. A lot of effort has been invested in trying to extract it quantitatively from the generalized parton distributions (GPDs) and the transverse-momentum dependent parton distributions (TMDs), which are accessed in high-energy processes and provide three-dimensional pictures of the nucleon. Recently, we have shown that it is more natural to access the quark OAM from the phase-space or Wigner distributions. We discuss the concept of Wigner distributions in the context of quantum field theory and show how they are related to the GPDs and the TMDs. We summarize the different definitions discussed in the literature for the quark OAM and show how they can in principle be extracted from the Wigner distributions.
\end{abstract}

\maketitle

\section{Introduction}	

The description of certain high-energy processes (like \emph{e.g.} deep-inelastic scattering) can be factorized into a process-dependent but perturbative part and a process-independent but non-perturbative part. The latter part contains crucial information about the internal structure of hadrons and, more generally, about the non-perturbative regime of quantum chromodynamics (QCD). In particular, one can access to different parts of the parton distributions in phase-space and study numerous correlations among the various degrees of freedom.

A unified picture of all this information is encoded in the so-called quantum phase-space or Wigner distributions. In the context of QCD, these distributions were first explored in Refs.~\cite{Ji:2003ak,Belitsky:2003nz} where relativistic effects were neglected. Recently, making the link with the generalized transverse-momentum dependent parton distributions (GTMDs) \cite{Meissner:2009ww}, we proposed a more detailed study which is not plagued by relativistic corrections \cite{Lorce:2011kd}. In particular, even though the Wigner distributions do not have a strict probablisitic interpretation due to the uncertainty principle, we showed that they can often be interpreted with semiclassical (and therefore more intuitive) pictures.

In this contribution, we briefly report on the phenomenology of the quark Wigner distributions obtained from relativistic light-front quark models \cite{Lorce:2011dv}, since so far no process is known for accessing directly to these distributions in experiments. We discuss in particular the relation between the Wigner distributions and the quark orbital angular momentum (OAM) \cite{Lorce:2011ni,Lorce:2011kn,Lorce:2012rr,Lorce:2012ce}.

\section{Wigner distributions}

As discussed in Ref. \cite{Lorce:2011kd}, using the impact-parameter representation \cite{Soper:1976jc,Burkardt:2000za,Burkardt:2002hr} the Wigner distributions are simply obtained \emph{via} two-dimensional Fourier transform
\begin{equation}\label{wigner}
\rho^{[\Gamma]q}_{\Lambda'\Lambda}(x,\vec k_\perp,\vec b_\perp,n)\equiv\int\frac{\ud^2\Delta_\perp}{(2\pi)^2}\,e^{-i\vec\Delta_\perp\cdot\vec b_\perp}\,W^{[\Gamma]q}_{\Lambda'\Lambda}(P,x,\vec k_\perp,\Delta,n).
\end{equation}
of the correlator \cite{Meissner:2009ww}
\begin{equation}\label{GTMDcorr}
W^{[\Gamma]q}_{\Lambda'\Lambda}(P,x,\vec k_\perp,\Delta,n)=\frac{1}{2}\int\frac{\ud^4z}{(2\pi)^3}\,\delta(z^+)\,e^{ik\cdot z}\,\langle p',\Lambda'|\overline\psi(-\tfrac{z}{2})\Gamma\,\mathcal W\,\psi(\tfrac{z}{2})|p,\Lambda\rangle,
\end{equation}
where $\Lambda$ and $\Lambda'$ are the initial and final hadron light-front helicities, respectively, the average hadron and quark four-momenta $P=(p'+p)/2$ and $k$, respectively, and the four-momentum transfer to the hadron $\Delta=p'-p$. We used for a generic four-vector $a^\mu=[a^+,a^-,\vec a_\perp]$ the light-front components $a^\pm=(a^0\pm a^3)/\sqrt{2}$. The variables $x=k^+/P^+$ and $\vec k_\perp$ are the average fraction of longitudinal momentum and average transverse momentum of the quark, respectively. The superscript $\Gamma$ is a matrix in Dirac space and $\mathcal W\equiv\mathcal W(-\tfrac{z}{2},\tfrac{z}{2}|n)$ is a Wilson line which ensures the color gauge invariance. The correlator \eqref{GTMDcorr} is parametrized in terms of the GTMDs~\cite{Meissner:2009ww}. They are very useful from a theoretical point of view, but only particular sections and projections of these functions are known to be accessible experimentally. For more details, see Refs.~\cite{Meissner:2009ww,Lorce:2011dv}.

\begin{figure}[h]
  \begin{minipage}[b]{0.5\linewidth}
   \centering
   \includegraphics[height=5cm]{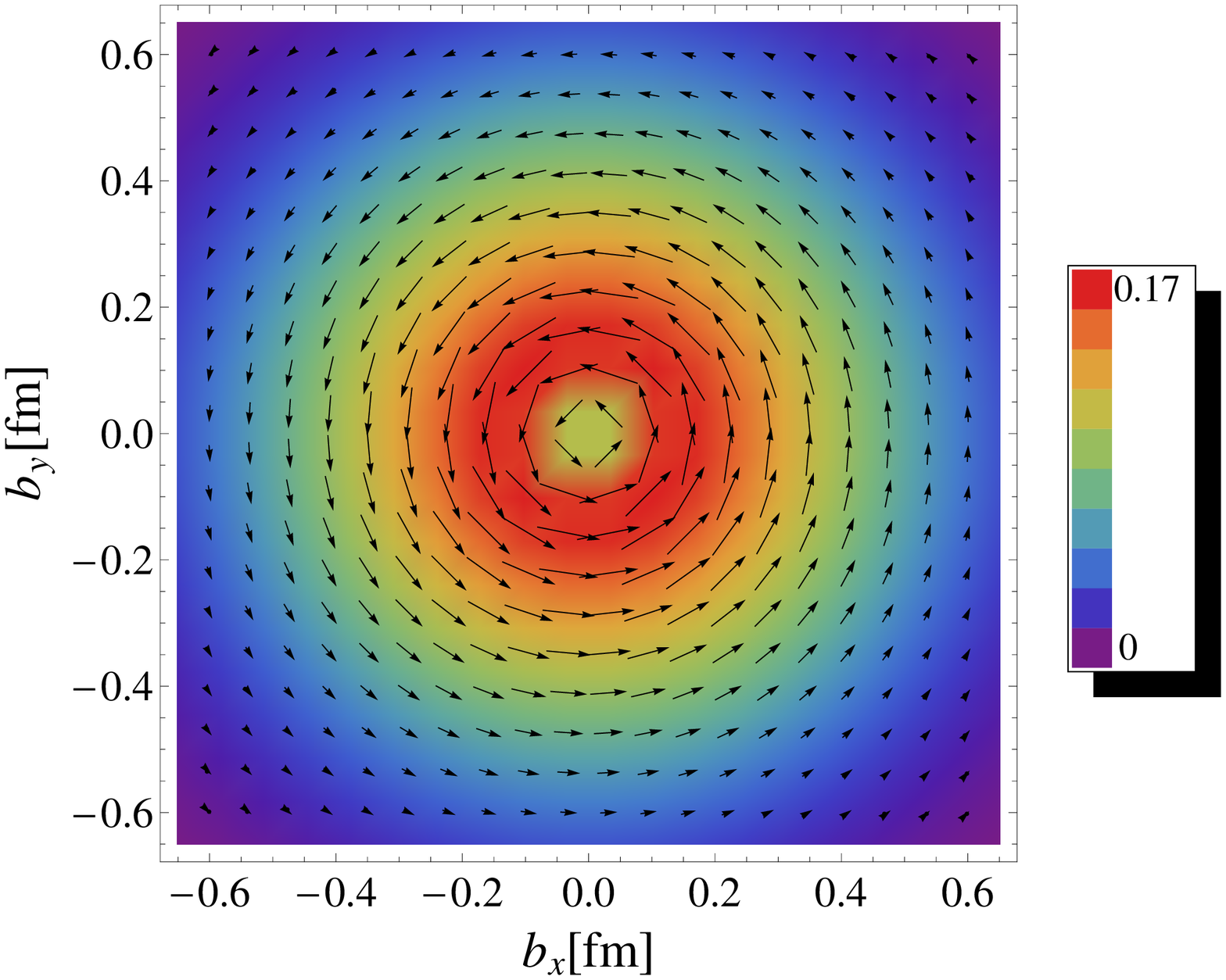}  
  \end{minipage}
  \begin{minipage}[b]{0.5\linewidth}
   \centering
   \includegraphics[height=5cm]{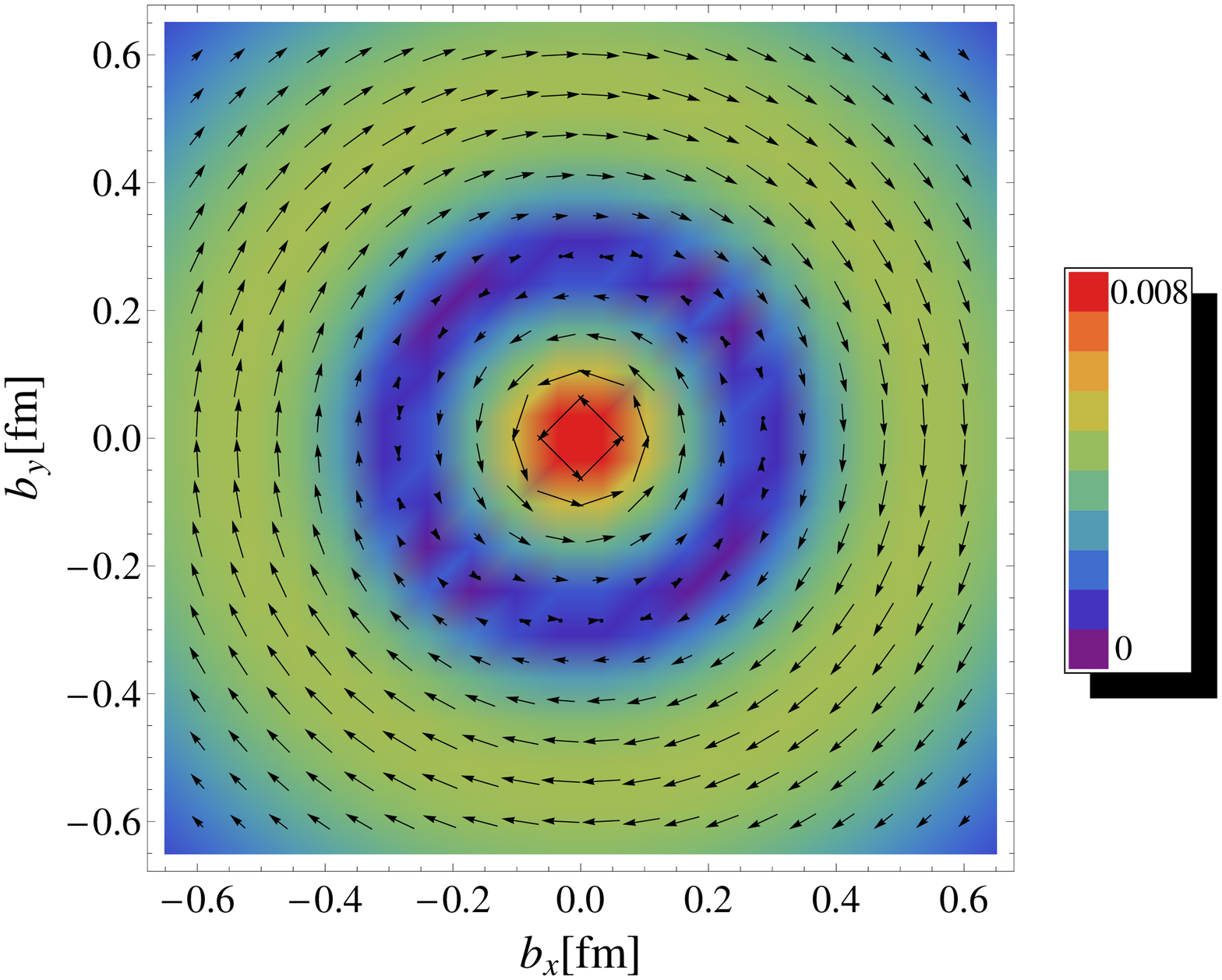}    
  \end{minipage}
\caption{Distributions in impact-parameter space of the average transverse momentum of unpolarized quarks in a longitudinally polarized nucleon. The nucleon polarization is pointing out of the plane, while the arrows show the size and direction of the average transverse momentum of the quarks. The left (right) panel shows the results for $u$ ($d$) quarks.}\label{fig1}
\end{figure}

At twist-two level, there exist 16 Wigner distributions. In the present contribution, we focus on the distribution of unpolarized quarks in a longitudinally polarized nucleon, which is directly related to the quark OAM. Other configurations for the quark and proton polarizations can be found in Ref.~\cite{Lorce:2011kd}. In Fig.~\ref{fig1} we show the impact-parameter distribution of the average quark transverse momentum in a longitudinally polarized nucleon \cite{Lorce:2011ni}
\begin{equation}
 \langle\vec k_\perp\rangle^q(\vec b_\perp)=\int\ud x\,\ud^2k_\perp\,\vec k_\perp\,\rho^{[\gamma^+]q}_{\Lambda\Lambda}(x,\vec k_\perp,\vec b_\perp,n).
\end{equation}
These figures have been obtained in the light-front constituent quark model (LFCQM) \cite{Boffi:2007yc,Pasquini:2007xz,Pasquini:2008ax,Boffi:2009sh,Pasquini:2010af,Pasquini:2011tk}, and are very similar to the ones obtained in the light-front version of the chiral quark-soliton model (LF$\chi$QSM) \cite{Lorce:2006nq,Lorce:2007as,Lorce:2007fa}. We clearly see that $u$ quarks have a net positive OAM, while $d$ quarks have globally a net negative OAM. Interestingly, we observe that the $d$-quark OAM changes sign around $0.25$ fm away from the transverse center of momentum.

\section{Quark orbital angular momentum}

The Wigner distributions are pretty intuitive objects as they correspond to phase-space distributions in a semiclassical picture. In particular, any matrix element of a quark operator can be rewritten as a phase-space integral of the corresponding classical quantity weighted with the Wigner distribution. It is therefore not so surprising that the longitudinal component of the quark OAM can simply be expressed as \cite{Lorce:2011kd}
\begin{equation}\label{OAMWigner}
l_z^q=\int\ud x\,\ud^2k_\perp\,\ud^2b_\perp\left(\vec b_\perp\times\vec k_\perp\right)_z\,\rho^{[\gamma^+]q}_{\Lambda\Lambda}(x,\vec k_\perp,\vec b_\perp,n).
\end{equation}
Since the Wigner distribution involves in its definition a gauge link, it inherits a path dependence. In Refs.~\cite{Ji:2012sj,Ji:2012ba} the authors claim that, when one chooses a straight gauge link, Eq.~\eqref{OAMWigner} simply gives the \emph{kinetic} OAM, \emph{i.e.} the one associated with the quark operator $-\frac{i}{2}\int\ud^3r\,\overline\psi^q\gamma^+(\vec r\times\!\stackrel{\leftrightarrow}{D}_r)_z\psi^q$, where $D_\mu=\partial_\mu-igA_\mu$ is the usual covariant derivative. It is therefore expected to give the same numerical result as the OAM obtained from the Ji relation \cite{Ji:1996ek} involving the generalized parton distributions
\begin{equation}\label{ji-sumrule}
L^q_z=\frac{1}{2}\int^1_{-1}\ud x\left\{x\left[H^q(x,0,0)+E^q(x,0,0)\right]-\tilde H^q(x,0,0)\right\}.
\end{equation}
However, a careful treatment of the gauge link indicates that this statement cannot be true \cite{Lorce:2012ce}. Instead, Eq. \eqref{OAMWigner} appears to be always related to the \emph{canonical} quark OAM, \emph{i.e.} the one associated with the quark operator $-\frac{i}{2}\int\ud^3r\,\overline\psi^q\gamma^+(\vec r\times\!\stackrel{\leftrightarrow}{\partial}_r)_z\psi^q$. In particular, straight gauge links give the gauge-invariant extension of the canonical OAM in the Fock-Schwinger gauge, while gauge links running along the light-front direction give the gauge-invariant extension of the canonical OAM in the light-front gauge. Recently, it has also been suggested that the TMD $h^\perp_{1T}$ may be related to the quark canonical OAM \cite{She:2009jq,Avakian:2010br}
\begin{equation}
\mathcal L_z^q=-\int\ud x\,\ud^2k_\perp\,\frac{k_\perp^2}{2M^2}\,h_{1T}^{\perp q}(x,k^2_\perp).
\end{equation}
However, this relation holds only within some model assumptions and no rigorous expression for the OAM in terms of the TMDs is known so far \cite{Lorce:2011kn}. For a more detailed discussion on the different decompositions and the corresponding OAM, see Ref.~\cite{Lorce:2012rr}.

\section{Conclusion}

In summary, we briefly introduced the quark Wigner distributions which can be interpreted in semi-classical terms as phase-space densities. We discussed how these Wigner distributions are simply related to the quark OAM, and showed the results within a relativistic light-front quark model calculation. Furthermore, we emphasized the path dependence due to the Wilson line entering the definition of the OAM in terms of Wigner distributions. In particular, we stressed that this definition corresponds always to a gauge-invariant extension of the canonical OAM, and has therefore no relation with the kinetic OAM.

\begin{theacknowledgments}
C. Lorc\'e is thankful to INFN and the Department of Physics of the University of Pavia for their hospitality. This work was supported in part by the Research Infrastructure Integrating Activity ``Study of Strongly Interacting Matter'' (acronym HadronPhysic3, Grant Agreement n. 283286) under the Seventh Framework Programme of the European Community, by the Italian MIUR through the PRIN 2008EKLACK ``Structure of the nucleon: transverse momentum, transverse spin and orbital angular momentum'', and by the P2I (``Physique des deux Infinis'') network.
\end{theacknowledgments}

\end{document}